\documentstyle[aps,prb,multicol,epsf]{revtex}
\begin{document}
\draft
\title{Quantum interference between multiple impurities in anisotropic superconductors}
\author{Brian M\o ller Andersen and Per Hedeg\aa rd}
\address{\O rsted Laboratory, Niels Bohr Institute for APG,
Universitetsparken 5, DK-2100 Copenhagen \O, Denmark}
\date{\today}
\maketitle
\begin{abstract}
We perform a numerical study of the quantum interference between impurities in d-wave
superconductors within a potential scattering formalism that
easily applies to multiple impurities. The evolution of the
low-energy local density of states for both magnetic and
nonmagnetic point scatterers is studied as a
function of the spatial configuration of the impurities. Further
we discuss the influence of a subdominant bulk superconducting order
parameters on the interference pattern from multiple impurities.
\end{abstract}
\begin{figure}
\centerline{\epsfxsize=\linewidth\epsfbox{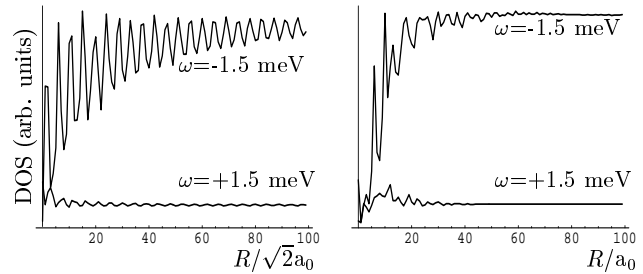}}
\caption{\label{spatialLDOS}DOS at (0,0) and at the
single-impurity resonance energy $\pm1.5$meV as a function of
distance between the two nonmagnetic impurities separated along the (a) nodal
direction and (b) anti-nodal direction. The y-axis scale is
identical for the two figures.}
\end{figure}
\begin{figure}
\centerline{\epsfxsize=\linewidth\epsfbox{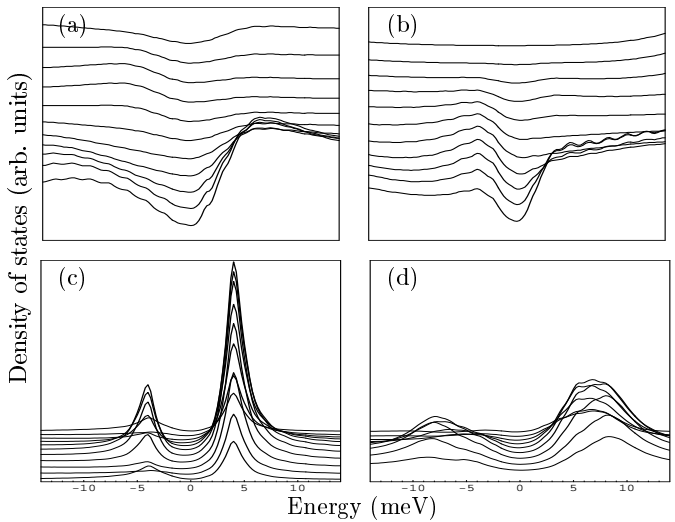}}
\caption{\label{LDOSscansMoveSTM}Low bias STM scans along the horizontal axis in
  steps of $a_0/5$ from $(0,0)$ (top line) to $(2,0)$ (bottom line). The scans
  are off-set for clarity. In (a) and (c)
there are two nonmagnetic impurities fixed at: (a) $(0,0)$ and $(1,0)$; (c) $(0,0)$
and $(2,0)$. In (b) and (d) there are three nonmagnetic impurities at: (b)
$(-1,0)$, $(0,0)$, $(1,0)$; (d) $(-2,0)$, $(0,0)$, $(2,0)$.}
\end{figure}
\begin{figure}
\centerline{\epsfxsize=\linewidth\epsfbox{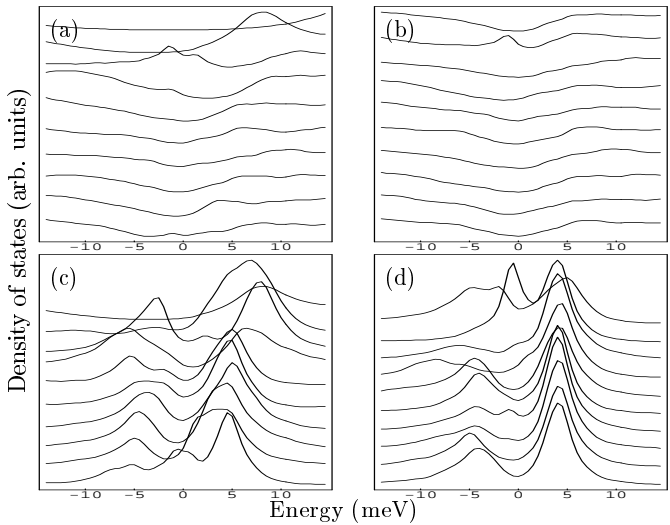}}
\caption{\label{m12diavert}LDOS at $(0,0)$ as a function of distance
  to a third impurity along the antinodal (a,c) or nodal (b,d)
  direction. The two fixed impurities are positioned at: In (a) and
  (b): $(-1,0)$ and $(0,0)$; in (c) and (d): $(-2,0)$ and $(0,0)$. In
  (a),(c) the third impurity is positioned at (top to bottom)
  $(1,0)$,$(2,0)$,...,$(10,0)$.In
  (b),(d) the third impurity is positioned at (top to bottom)
  $(1,1)$,$(2,2)$,...,$(10,10)$.}
\end{figure}
\begin{figure}
\centerline{\epsfxsize=\linewidth\epsfbox{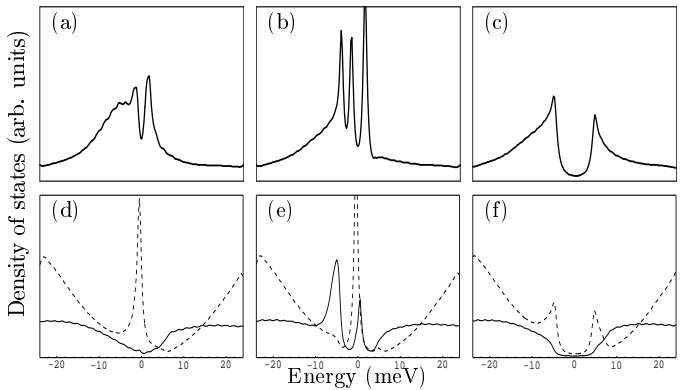}}
\caption{\label{2and3impsddxy}Top row: DOS at $(0,0)$ for two nonmagnetic impurities at
(0,0) and (2,4). Bottom row: DOS at $(0,0)$ (solid line) and $(1,1)$
(dashed line) for three nonmagnetic impurities at (-1,1), (1,-1) and (-1,-1). Pairing symmetry: (a) and (d) d$_{x^2-y^2}$,
(b) and (e) d$_{x^2-y^2}$+id$_{xy}$, (c) and (f) d$_{x^2-y^2}$+is.}
\end{figure}
\begin{figure}
\centerline{\epsfxsize=\linewidth\epsfbox{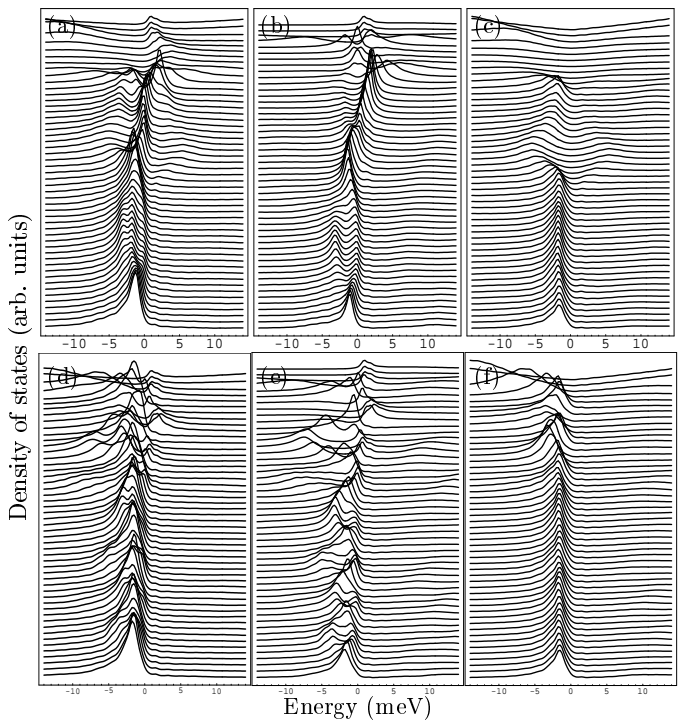}}
\caption{\label{magscahorz}DOS at $(0,0)$ for: (a,d) one magnetic and one nonmagnetic impurity,
(b,e) two magnetic ($V^M_1=V^M_2$), (c,f) two magnetic
($V^M_{1}=-V^M_{2}$). The topmost graph in each figure is the DOS when the two scatterers are both
positioned at the origin whereas in the lowermost at $(0,0)$ and
$(10,0)$. Antinodal separation: (a-c), and nodal separation: (d-f).}
\end{figure}

\begin{multicols}{2}
\noindent The past few years have proved the importance of
experimental techniques which can directly test the wealth
of information associated with modifications of the local
density of states (LDOS) around impurities, grain boundaries and
vortices in strongly correlated electron systems. In particular
scanning tunneling microscopy (STM) measurements have provided detailed LDOS images
around single nonmagnetic\cite{yazdani,pan} (Zn) and
magnetic\cite{hudson} (Ni) impurities on the surface of the
high temperature superconductor Bi$_2$Sr$_2$CaCuO$_{4+\delta}$ (BSCCO).\\
For conventional superconductors Yu and Shiba\cite{shiba} showed
that as a result of the interaction between a magnetic impurity
and the spin density of the conduction electrons, a bound state
located around the magnetic impurity is formed inside the gap in
the strong-scattering (unitary) limit. For anisotropic
superconductors a number of authors generalized the Yu-Shiba
approach to study the LDOS around single
impurities\cite{rosengren}. It was found, for instance, that for a
single nonmagnetic impurity there is only a virtual bound
(or resonant) state due to the existence of the
low-energy nodal quasi-particles. The one-impurity problem was
recently reviewed by several authors\cite{morrstravropoulos,hirschfeld}.\\
Recently Hoffman {\sl et al.}\cite{hoffman} measured the energy dependence of
the Fourier transformed LDOS images on the surface of optimally doped BSCCO below
T$_c$. The dispersive features were explained from
the point of elastic quasi-particle interference resulting from a single
{\sl weak}, nonmagnetic impurity\cite{dhlee}. This gives credence that
a scattering potential picture can yield valuable predictions in the
superconducting state of these materials. Evidence for quantum interference between
{\sl strong} scatterers has been observed in the CuO chains of
YBa$_2$Cu$_3$O$_{6+x}$ by Derro {\sl et al.}\cite{derro} Future experimental ability
to control the position of the impurities on the surface of a superconductor
and perform detailed STM measurements around multiple impurity
configurations motivates further theoretical studies of the resulting quantum interference.\\
Previous calculations have studied the formation of bonding and antibonding states around two
magnetic impurities in s-wave superconductors\cite{flattereynolds,morrstrav}.
For d-wave superconductors it was found that the interference effects between two nonmagnetic
unitary scatterers depends crucially on the distance and orientation of the
impurities\cite{morrstravropoulos,hirschfeld,onishi}. The
orientational dependence arises from the anisotropic gap function,
and provides an alternative method to identify the symmetry of the
superconducting gap. Several authors\cite{rosengren,flatte} have
previously suggested similar ideas in the case of one impurity.\\
In this paper we study multiple impurity effects by exactly
inverting the Gorkov-Dyson equation. In particular, we discuss the
effect of quantum interference between: 1) nonmagnetic impurities
in the strong scattering limit, 2) nonmagnetic
impurities in the case of induced subdominant superconducting
order parameters, and 3) magnetic and nonmagnetic impurities. All
the calculations are performed within a quasi-particle scattering
framework with classical impurities\cite{shiba,flatte}. The main
purpose is to use quantum interference to obtain results that pose
further tests on this approach and to illustrate the strong
sensitivity of the LDOS on the positions of the impurities.\\
The Greens function $\hat{G}^{(0)}\!({\mathbf{k}},\omega)$ for the unperturbed d-wave superconductor is
given in Nambu space by
\begin{equation}
\hat{G}^{(0)}\!({\mathbf{k}},\omega)\!=\!\left[ i\omega_n\hat{\tau}_0\!-\!\xi({\mathbf{k}})\hat{\tau}_3\!-\!\Delta({\mathbf{k}})\hat{\tau}_+
\!\!-\!\Delta^*\!({\mathbf{k}})\hat{\tau}_- \right]^{-1},
\end{equation}
where $\hat{\tau}_\nu$ denotes the Pauli matrices in Nambu space,
$\hat{\tau}_0$ being the $2\times2$ identity matrix and
$\hat{\tau}_\pm=\hat{\tau}_1 \pm \hat{\tau}_2$. For a system with
d$_{x^2-y^2}$-wave pairing $\Delta({\mathbf{k}})=\frac{\Delta_0}{2}
\left( \cos (k_x) - \cos (k_y) \right)$. Below,
$\Delta_0=25\mbox{meV}$ and the lattice constant $a_0$ is set to unity.
In this article we use a normal state quasi-particle energy
$\xi(\mathbf{k})$ relevant for BSCCO around optimal doping (14\%)
\begin{equation}
\xi({\mathbf{k}})\!=\!-2t \left( \cos (k_x)\! +\! \cos (k_y) \right)
\!-\!4t' \cos (k_x) \cos (k_y)\!-\!\mu
\end{equation}
with $t=300 \mbox{meV}$, $t'=-0.4t$ and $\mu=-1.18t$. Here $t$ $(t')$
refers to the nearest (next-nearest) neighbor hopping integral
and $\mu$ is the chemical potential.\\
We model the presence of scalar and magnetic impurities in the
system by the following $\delta$-function interactions
\begin{equation}
\hat{H}^{int}\!=\! \sum_{i} \left[ \left( V^S_i\!+\!V^M_i \right)
\hat{c}^\dagger_{i\uparrow} \hat{c}_{i\uparrow} +
\left( V^S_i\!-\!V^M_i \right) \hat{c}^\dagger_{i\downarrow}
\hat{c}_{i\downarrow} \right].
\end{equation}
Here $i$ denotes the set of lattice sites containing magnetic
and/or scalar impurities and $V^M_i$ ($V^S_i$) is the strength of
the corresponding effective potential. We consider only the
$z$-component of the magnetic impurity interaction and ignore
spin-flip scattering.\\
For a single nonmagnetic impurity at the origin it is well-known
that the scattering modifies the Greens function by
\begin{equation}\label{oneimpres}
\delta G_{11}({\mathbf{r}},i\omega_n)\!=\!\!\frac{\left[G^{(0)}_{11}({\mathbf{r}},i\omega_n)\right]^2}{\frac{1}{V^S}\!-\!G^{(0)}_{11}(0,
i\omega_n)}\!-\!\frac{\left[G^{(0)}_{12}({\mathbf{r}},
i\omega_n)\right]^2}{\frac{1}{V^S}\!-\!G^{(0)}_{11}(0, -i\omega_n)}.
\end{equation}
Here $r$ is the distance to the origin and $G_{\alpha\beta}$ refers to
the $\alpha\beta$th entry of the $2\times2$ Nambu subspace.\\ 
Naturally one can derive equivalent expressions for the LDOS
modulations around several impurities. However, for a numerical
study of the evolution of the LDOS for multiple impurities positioned
at arbitrary lattice sites, we find it is easier to invert directly the
real-space Gorkov-Dyson equation. The full Greens function $\hat{G}({\mathbf{r}},\omega)$ is then
obtained by solving the equation
\begin{equation}
\underline{\underline{\hat{G}}}(\omega) =
\underline{\underline{\hat{G}}}^{(0)}\!(\omega) \left(
\underline{\underline{\hat{I}}} -
\underline{\underline{\hat{H}}}^{int}
\underline{\underline{\hat{G}}}^{(0)}\!(\omega) \right)^{-1},
\end{equation}
where the double lines indicate that the elements of this
equation are matrices written in real- and Nambu space. The size
of these matrices depends on the number of impurities and the
dimension of the Nambu space. We have previously utilized this
method to study the electronic structure around vortices that
operate as pinning centers of surrounding
stripes\cite{andersenpinned}. We perform the 2D Fourier transform
of the clean Greens function $\hat{G}^{(0)}({\mathbf{k}},\omega)$
numerically by dividing the first Brillioun zone into a
$1400\times1400$ lattice
and introducing a quasi-particle energy broadening of $\delta = 0.5
\mbox{meV}$ with $\delta$ defined from $i\omega_n\rightarrow\omega+i\delta$.\\
The differential tunneling conductance is proportional to the LDOS
$\rho({\mathbf{r}},\omega)$ which in turn is determined from
\begin{equation}
\rho({\mathbf{r}},\omega)=-\frac{1}{\pi}\mbox{Im} \left[
G_{11}({\mathbf{r}},\omega) + G_{22}({\mathbf{r}},-\omega)
\right].
\end{equation}
In the following we model the nonmagnetic unitary scatterers with a potential
$V^S=700\mbox{meV}$ which gives rise to resonance energies around
$\pm1.5\mbox{meV}$ in agreement with experiment\cite{pan}. (This is
seen from the holelike resonance evident in the bottom LDOS scan in Fig.
\ref{magscahorz}a. For a single impurity only one of the two
resonances evident from Eqn. \ref{oneimpres} has weight on the
impurity site since the anomalous part of the Greens function,
$G_{12}(r,\omega)$, vanishes at $r=0$ due to the symmetry of the
  d-wave gap.)\\    
For interference between two nonmagnetic unitary impurities Morr
{\sl et al.}\cite{morrstravropoulos} found strong variations in
the LDOS as the distance between the impurities $R$ is varied
along one of the crystal axes.
The single-impurity spectrum was obtained when $R$ exceeds
approximately $10a_0$.
However, as expected for a $d_{x^2-y^2}$-wave
superconductor, this length scale is much larger along the nodal
directions. This is seen from Fig. \ref{spatialLDOS}. Here the
density of states is measured above one of the impurities fixed at the
origin while the other is moved away along the nodal (a) or anti-nodal
(b) direction. The single impurity LDOS is obtained for $R$ well
above $100a_0$. Thus only for impurity concentrations below 0.1\% does the LDOS
correspond to the expected result from a single strong nonmagnetic
impurity. For weaker scatterers the decay length will be considerably
reduced.
\noindent For quantum interference between multiple fixed 
nonmagnetic unitary scatterers Fig.
\ref{LDOSscansMoveSTM} shows the LDOS as the STM tip is scanned
along a crystal axis on which the impurities are positioned.
\noindent In general the resonances are split by
the proximity of other impurities and the number of resonances is
directly proportional to the number of interfering
impurities. However, locally the density of states may be strongly influenced by
destructive interference. For instance, for the two impurities (Fig.
\ref{LDOSscansMoveSTM}a,c) sharp resonances exist only when $R=2a_0$ as is evident from
Fig. \ref{LDOSscansMoveSTM}c. When a third impurity is added at $(-2,0)$ these resonances
appear to broaden and shift to higher energies,
Fig. \ref{LDOSscansMoveSTM}d. Contrary to this,
Fig. \ref{LDOSscansMoveSTM}a,b show that the
addition of a third impurity has only a minor effect when $R=a_0$.\\
The case of three nonmagnetic impurities is studied further in Fig.
\ref{m12diavert} which shows the evolution of the LDOS at $(0,0)$ as a
function of the distance to a third impurity along the nodal (b,d) and
anti-nodal (a,c) direction. The case without the third impurity corresponds to
the topmost LDOS in Fig. \ref{LDOSscansMoveSTM}a and \ref{LDOSscansMoveSTM}c.    
\noindent As in the case of two nonmagnetic impurities\cite{morrstravropoulos,hirschfeld}
there are very strong variations in the final LDOS; the number of
apparent resonances, their width and resonance energy depends
crucially on the positions of the three impurities. The small
modulations added by the third impurity seen in Fig. \ref{m12diavert}a,b
agree with the destructive interference evident from the
corresponding cases seen in Fig. \ref{LDOSscansMoveSTM}a,b. Contrary
to this, large modulations are again seen when increasing the distance
between the two fixed impurities by a single lattice constant, Fig. \ref{m12diavert}c,d.\\
Recently Zhu {\sl et al.}\cite{hirschfeld} suggested a careful study of the
two-impurity problem to extract information of the bulk Greens
function of the clean system. In the following we show how the quantum interference between unitary
scatterers is strongly affected by the induction of a small
subdominant superconducting order parameter. Thus one may 
utilize the quantum interference between several impurities as an 
alternative method to detect a small subdominant order parameter.\\ 
For instance, tuning
through a quantum phase transition from a d$_{x^2-y^2}$ to
a d$_{x^2-y^2}$+id$_{xy}$ superconductor at a critical doping
level\cite{vojta}, magnetic impurity concentration\cite{mb} or
magnetic field strength\cite{laughlin}, a small d$_{xy}$ order
could {\sl qualitatively} alter the interference pattern. For
$\Delta_{xy}({\mathbf{k}})=\Delta_{xy}^0 \sin (k_x) \sin (k_y)$
with $\Delta_{xy}^0$=5.0meV, we compare in Fig.
\ref{2and3impsddxy} the LDOS at physically realizable impurity
positions to the case with $\Delta_{xy}^0$=0.
\noindent Also we show the
difference between $d+id$ and $d+is$ pairing symmetry for these
impurity configurations. For most spatial configurations the secondary
pairing (id or is) leads to a sharpening of the resonances but at particular
positions there is a qualitative difference as shown in
Fig. \ref{2and3impsddxy}. For instance, the induction of d+id pairing
(Fig. \ref{2and3impsddxy}b) can result in three apparent resonances
contrary to the ground state with pure
d$_{x^2-y^2}$-wave pairing (Fig. \ref{2and3impsddxy}a). Similarly, by
comparing the LDOS at (1,1) (dashed lines) in
Fig. \ref{2and3impsddxy}d-f, it is evident that the interfering
scatterers can provide a clear distinction between d+id and d+is pairing.
Information of the induction of {\sl
  local} order around the impurities can also be inferred from
STM measurements of specific impurity configurations\cite{mb}.\\
We turn now briefly to the study of the classical magnetic
impurities in d$_{x^2-y^2}$-wave superconductors. The interference
of two magnetic scatterers in a s-wave superconductor was
studied recently by Flatte {\sl et al.}\cite{flattereynolds} For comparison
to the nonmagnetic case we use a magnetic potential strength
$|V^M|=700\mbox{meV}$ which does, however, not model all magnetic impurities (e.g. Ni) on the
surface of BSCCO\cite{hudson,flatteand}. Future
experiments will reveal whether the scattering potential formalism
utilized here is appropriate or whether more correlated
effects are required\cite{wangplee,polkovnikov}.\\
Fig. \ref{magscahorz} shows the quantum interference
between two unitary scatterers: (a,d) one magnetic and one
nonmagnetic, (b,e) two parallel magnetic, and (c,f) two antiparallal
magnetic. In all figures one impurity is fixed below the STM tip at
the origin $(0,0)$ while the other is displaced along the: (a-c)
horizontal crystal axis or (d-f) along the nodal direction.
In Fig. \ref{magscahorz}a,d it is the nonmagnetic impurity that
is fixed at the origin.
Again the number of resonances, their position, amplitude and
width depends in detail on the distance and nature of the two
scatterers. Further, the spatial evolution of the LDOS is quite similar for
case (a),(b), and (d),(e). These are, however, very different from the interference between
two antiparallel magnetic impurities (c,f) which is dominated by strong
destructive interference at small separations along the antinodal
direction and a surprisingly fast recovery to the single
impurity case along both the nodal and antinodal directions.\\
\noindent The results presented above remain
qualitative since fits to a specific experiment in addition to details
from the tunneling matrix elements could also include
modified hopping integrals around the impurities, gap suppression and
possibly {\sl both} magnetic and nonmagnetic scattering potentials\cite{flatteand}. We
have checked that a gap suppression on the bonds around the impurity
site does not produce any qualitative changes\cite{dhlee}. However, on a
phenomenological level the gap suppression could allow for a competing
magnetic order parameter to develop around the impurity. Thus, the gap
suppression might be important for
explaining the formation of magnetic moments around nonmagnetic
impurities as seen by NMR experiments. These
issues are currently controversial but the vast amount of information
inferred from the quantum interference between multiple impurities may
help settle this, and more importantly settle the validity of the
scattering potential scenario versus more correlated models.

\end{multicols}
\end{document}